\documentclass[12pt]{article}
\usepackage{graphicx}
\usepackage{amsmath,amssymb,xcolor}

\newcommand\pubnumber{TTK-10-57}
\newcommand\pubdate{\today}

\def\napoli{Institut f\"ur theoretische Teilchenphysik und Kosmologie\\
RWTH Aachen, D-52074 Aachen, GERMANY}

\def\Title#1{\begin{center} {\Large #1 } \end{center}}
\def\Author#1{\begin{center}{ \sc #1} \end{center}}
\def\Address#1{\begin{center}{ \it #1} \end{center}}

\newcommand\pubblock{\rightline{\begin{tabular}{l} \pubnumber\\
         \pubdate  \end{tabular}}}
\newenvironment{Abstract}{\begin{quotation}  }{\end{quotation}}
\newenvironment{Presented}{\begin{quotation} \begin{center} 
             PRESENTED AT\end{center}\bigskip 
      \begin{center}\begin{large}}{\end{large}\end{center} \end{quotation}}
\def\Acknowledgements{\bigskip  \bigskip \begin{center} \begin{large}
             \bf ACKNOWLEDGEMENTS \end{large}\end{center}}

\begin{document}
\begin{titlepage}
\pubblock

\vfill
\Title{On the Status of $V_{td}$, $V_{ts}$ and $V_{tb}$}
\vfill
\Author{ J\"urgen Rohrwild}
\Address{\napoli}
\vfill
\begin{Abstract}
The CKM matrix elements $V_{td}$, $V_{ts}$ and $V_{tb}$, while notoriously hard 
to measure directly in experiment, can be determined to great precision
within the Standard Model (SM) with a minimal set of observables due to unitarity 
of $V_{\rm CKM}$. 
We discuss the potential modifications of these elements in extensions with
non-unitary quark mixing matrices. 
\end{Abstract}
\vfill\begin{center} \large
Proceedings of CKM2010, the 6th International Workshop on the CKM Unitarity Triangle 
\end{center}
\vfill
\begin{Presented}
     University of Warwick, UK, 6-10 September 2010
\end{Presented}
\vfill
\end{titlepage}
\def\thefootnote{\fnsymbol{footnote}}
\setcounter{footnote}{0}

\section{Introduction}
The precise determination of CKM matrix elements---be it their absolute value
or the relative phases---and the verification of CP violation via the CKM mechanism 
\cite{Kobayashi:1973fv} were milestones of particle physics in the last decade(s).
Leptonic and semi-leptonic meson decays as well as superallowed $\beta$ decay have 
provided an excellent idea about the absolute values of the element of the
first two rows of the CKM matrix.  

Unfortunately, direct measurements of the {\it'top'}-row elements cannot be carried out 
in similar fashion as the $t$ width is too large for it to hadronize. Hence, direct information 
on these elements is fairly rare. On the other hand, a large amount of loop-dominated observables 
are very sensitive to the $V_{tx}$ elements, e.g., $B_d$ mixing, $B_s$ mixing, or $b\to s \gamma$,
and thus provide indirect information on these elements. We use ``indirect''
to emphasise that one usually makes use of some Standard Model (SM) properties (e.g. CKM unitarity) 
to extract the desired information. 

This also raises the question to what extent a non-standard CKM matrix arising in the low energy 
limit from some beyond the SM (BSM) scenario can differ from the SM form, if one takes into account the 
large amount of precision studies dedicated to the determination of the CKM matrix. In the following 
I will discuss the high precision provided by the unitarity condition in the SM and some BSM scenarios 
which lead to notable deviations from the SM values for the $V_{tx}$ elements.

\section{Experiment and Standard Model}

Before examining the SM predictions for the {\it top} row matrix elements, let us first look 
at the few direct experimental constraints. There are, in principle, two ways to directly constrain 
the matrix element $V_{tb}$\footnote{For specific BSM scenarios other options 
may exist.}: $t \overline t$ and single top production.
\begin{itemize}
 \item  $t \overline t$ production allows for a measurement of the ratio $\mathcal{R}_b$
which is defined as 
 \begin{align}
       \mathcal R _b =\frac{\mathcal{B}[t\to Wb]}{\mathcal{B}[t\to Wq]}=
               \frac{|V_{tb}|^2}{|V_{td}|^2 + |V_{ts}|^2 + |V_{tb}|^2}\;. 
  \end{align}
D\O{} determined $\mathcal{R}_b$ to be  $0.97^{+0.09}_{-0.08}$ \cite{Abazov:2008yn}. This immediately 
translates to a value for $V_{tb}=0.97^{+0.09}_{-0.08}$ in the SM. However, the model-independent conclusion 
is merely $V_{tb}\gg V_{td},V_{ts}$ as one cannot use unitarity.
 \item The single top production cross section is directly proportional to $|V_{tb}|^2$. The combined CDF- D\O{} 
result \cite{Group:2009qk} (assuming a top quark mass of $170\;\rm GeV$) leads to $|V_{tb}|=0.88\pm 0.07\pm 0.07 \;({\rm exp}+{\rm theo})$
 (see \cite{Aaltonen:2010jr} for a recent CDF update and Wolfgang Wagner's talk \cite{ckm2010} for plots and details). 
Note that LHC can reduce the experimental uncertainty by almost a factor of 2 \cite{Aad:2009wy, Ba2007, Han:2008xb}.
\end{itemize}

From the SM point of view the issue of $V_{td}$, $V_{ts}$ and $V_{tb}$ is of course already settled.
The easiest way to determine not only the absolute values but also relative phases is the
use of CKM unitarity which reduces the number of independent parameters to 4. In fact one can even avoid
using any 'loop-related' observables by only making use of the absolute values of 
$|V_{ud}|$, $|V_{us}|$, $|V_{ub}|$, $|V_{cd}|$, $|V_{cb}|$ and  $|V_{cs}|$ from 
tree-level decays and by supplying a single phase via the CKM angle $\gamma$ from e.g. $B\to D^{(*)}K$ which is tree-level dominated. 
The parameters of the  CKM matrix can then be fitted; the UTfit group e.g. obtains \cite{UTfit}: \\
{\tiny
\begin{center}
$$\!\!\!\!V_{CKM}\! \! = \!\begin{pmatrix}
             0.9426\pm 0.00015 & \!\!\!\!\!0.22535\pm 0.00065 &  
                   \!\!\!\!\!\!\!\!0.00376 \pm 0.0002\!\cdot\! e^{i(-73.8\pm9.4)^\circ}\\
            -0.2252\pm 0.00065 \!\cdot\! e^{i(-0.03656\pm 0.0028)^{\circ}} 
                 & \!\!\!\!\!0.97345\pm 0.00015 & \!\!\!\!\!\!\!\! 0.04083\pm 0.00045\\
           0.00896\pm\;0.0006\;\;{\&}\;\; 0.01081\!\pm\! 0.0006\!\cdot\!e^{i(-22.9\pm 1.4)^\circ}&
               -0.03979\pm\! 0.00052\!\cdot\! e^{i(-1.163\pm 0.084)^\circ}&  
              0.99916\pm\! 1.8\!\times\!10^{-5}
\end{pmatrix}$$
\end{center}}
With exception of some ambiguity in $V_{td}$ the values of $V_{ts}$ and $V_{tb}$ are already
determined with astounding accuracy. However, one can do even better by including 
the plethora of precision flavour measurements in the analysis. This provides a powerful
consistency check for unitarity and results in the famous unitarity triangle plots, see 
e.g.~\cite{Charles:2004jd}.

\section{Beyond Unitarity}
 As we have seen in the previous section unitarity of the CKM matrix already
pins down the magnitude of the $V_{tx}$ elements to great accuracy even if 
conservative input data is used. Hence, a deviation from the SM values
basically requires a breaking of the unitary condition. The most general
parameterisation of the CKM matrix then requires 13 independent parameters,
see e.g.~\cite{Kim:2000gv}. Of course, it is not meaningful to ask for a 
general determination of this parameters as  non-unitarity {\it must}
be induced by some physics beyond the SM, whose effects would have to be 
taken into account. 

Most scenarios generate the deviation of the CKM matrix 
by enlarging the fermion sector. Thus the SM CKM matrix is a $3\times 3$
subblock embedded in a larger fermion mixing matrix. While there are 
numerous BSM scenarios, this talk will focus only on two in a sense minimal 
models and briefly discuss a third, more involved extension.

\subsection{Additional Fermions}

The simplest way to break unitarity of the SM CKM matrix is to either
introduce one heavy vector-like quark or a complete additional (SM-like) 
fermion generation. In both cases the experimental data must, strictly speaking, 
be reinterpreted as theory input cannot make use of any identity like 
  $$ V_{td} V_{tb}^* + V_{cd} V_{cb}^* + V_{ud} V_{ub}^*=0\;. $$
Furthermore, the possibility of modifications to other well-known features of 
the SM flavour structure cannot be ruled out a priori.

\paragraph{Vector-like quarks} generically make an appearance in various extensions of the Standard
 model like Randall-Sundrum scenarios or $E_6$ GUTs. The minimal 
formulation just extends the fermion sector by a single heavy top or bottom like quark, see
 e.g.~\cite{Kim:2007zzg, Botella:2008qm}. Restricting to the case of a charge $+2/3e$ vector 
quark $T$ one finds the following features:
\begin{itemize}
 \item The equivalent of the CKM matrix is now $4\times 3$
    \begin{align}\label{CKMvec}
     V_{CKM} = \begin{pmatrix}
                    V_{ud}&V_{us}&V_{ub}\\
                    V_{cd}&V_{cs}&V_{cb}\\ 
                    V_{td}&V_{ts}&V_{tb}\\ \hline
                    V_{Td}&V_{Ts}&V_{Tb}
                   \end{pmatrix}\;.
    \end{align}
\item As the Yukawa matrix is no longer diagonal in the mass eigenbasis the Higgs interaction is flavour changing.
\item The $Z$ can induce flavour changing neutral currents (FCNCs). 
      Their strength is proportional to the unitarity violation \cite{Botella:2008qm}.
\item The heavy $T$ will preferable mix with the top quark; thus reproducing the 
      unitary relations of first and second row within experimental accuracy.
\end{itemize}

The entries of the matrix in Eq.~\eqref{CKMvec} and value of the mass of the $T$ 
are subject to a large amount of experimental constraints coming from:
direct measurement of CKM elements, meson mixing 
($\varepsilon_K, \varepsilon'/\varepsilon_K$, mass differences in the $B_{d/s}$ systems,\ldots ), 
various rare kaon and $B$ decays as well as bounds from 
from the electroweak sector such as $S$ and $T$ parameter and the ratio $R_b$
\cite{Alwall:2006bx}.
The two parameter sets below (taken from \cite{Botella:2008qm}) illustrate 
the size of the modification the CKM matrix (only absolute values of the 
elements are shown) can still experience:\vspace{0.3cm}\\
{\scriptsize
\begin{minipage}{0.48\textwidth}
  {\scriptsize $m_{T}=450\;\rm GeV$}\vspace{-0.1cm} \hfill 
 $$|U_{D}|=\left(\begin{matrix}0.974179& 0.225657& 0.004031\ \\ 0.225619& 0.972525& 0.041766\ \\
 \colorbox{red!20}{$0.008330$}&\colorbox{red!20}{$ 0.047219$}& \colorbox{red!20}{$0.966377$}\ \\ 0.001136& 0.032304& 0.253683\ \end{matrix}\right) $$
\end{minipage}$\;\;\;\;$
\begin{minipage}{0.48\textwidth}
   {\scriptsize $m_{T}=300\;\rm GeV$}\vspace{-0.1cm} \hfill 
 $$|U_{D}|=\left(\begin{matrix}0.974195& 0.225663& 0.004137 \\ 0.2254882& 0.972938& 0.041548 \\
 \colorbox{red!20}{$\!0.009721\!$}&\colorbox{red!20}{$\! 0.042034\!$}& \colorbox{red!20}{$\!0.945531\!$}\ \\
 0.002889& 0.026471& 0.322842 \end{matrix}\right) $$
\end{minipage}\footnotesize\vspace{0.2cm}\\ }

\paragraph{One additional fermion generation:}

\begin{figure}
\begin{minipage}{0.32\textwidth}
\begin{center}
 \includegraphics[width=0.97\textwidth]{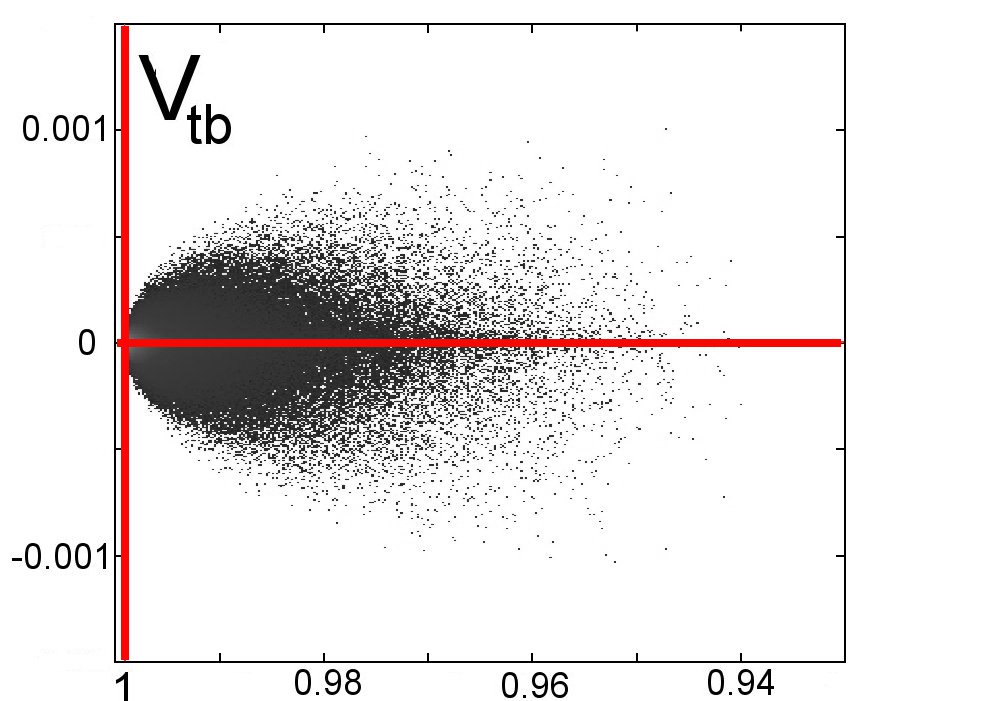}
\end{center}
\end{minipage}
\begin{minipage}{0.32\textwidth}
\begin{center}
 \includegraphics[width=0.97\textwidth]{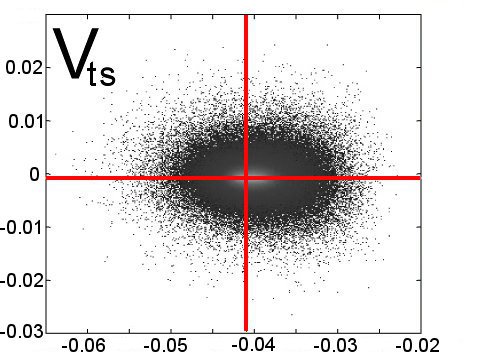}
\end{center}
\end{minipage}
\begin{minipage}{0.32\textwidth}
\begin{center}
 \includegraphics[width=0.97\textwidth]{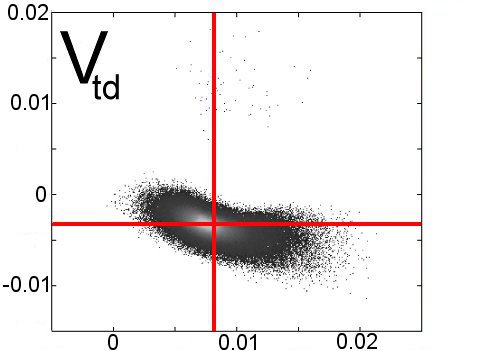}
\end{center}
\end{minipage}
 \caption{Scatterplots for $V_{tb}$, $V_{ts}$ and $V_{td}$ (standard representation) in the complex plane 
for the 4G scenario \cite{cancellations}. The crossed lines denote the SM values.\label{f1}}
\end{figure}

Introducing a SM-like generation (adding 7 new parameters in the quark sector)
is conceptionally even simpler then the vectorlike quark model. However, 
the absence of FCNCs and gauge anomalies comes with a massive fourth neutrino\footnote{The impact of 
the large mass and the implications for the PMNS matrix are discussed e.g.~in \cite{neutrinos}.}. Both
 CKM and PMNS matrices have to be promoted to $4\times 4$ matrices; the SM CKM matrix is a sub block of 
the unitary matrix 
$$V_{CKM4}=\left(
\begin{array}{ccc|c}
V_{ud} & V_{us} & V_{ub} & V_{ub'}\\
V_{cd} & V_{cs} & V_{cb} &   V_{cb'}\\
\colorbox{red!20}{$V_{td}$} & \colorbox{red!20}{$V_{ts}$} & \colorbox{red!20}{$V_{tb}$} & V_{tb'}\\ \hline
V_{t'd} & V_{t's} & V_{t'b} & V_{t'b'}\\
\end{array}
\right)\;.$$ 
Important bounds on the possible values of $V_{td}$, $V_{ts}$ and $V_{tb}$
come from the same $\Delta F =1,2$ processes already mentioned in the vector quark scenario, see \cite{Flavourfourth}. However, 
the strongest constraints are indeed provided by electroweak precision observables \cite{Eweakfourth}; e.g.~the renowned $S$ 
and $T$ parameters provide strong limits on the mixing of fourth and third generation.
Still, CKM elements can be modified by quite a bit compared to the SM. Fig.~\ref{f1} illustrates this for the
third row elements.
Especially the ``small'' elements $V_{ts}$ and $V_{td}$ can receive large relative modifications.
Note that without electroweak constraints $V_{tb}$ could be as low as 0.8 and still survive the 
bounds set by flavour physics alone.
 
Both, vector quark and extra generation, allow deviations of roughly the same large size.
At first glance this result is rather surprizing as one would expect loop-dominated flavour
observables to be very sensitive to both, new heavy particles ($T$ or $t'$) in loops
and modified $W$ couplings due to the changed CKM element values, 
 and thus veto any drastic modifications. However, the two types of contributions
(which may be supplemented by small tree-level FCNCs for the vector quark scenario)
 can cancel each other to a rather large extend. In fact this does not seem unnatural 
 as the SM CKM structure is 'thinned' out in order to accommodate for $4\times 4$ 
 unitarity, while an additional contribution due to the new fermions has to be added.
 This behaviour is observed for almost all parameter sets that result in large 
 modifications of the CKM elements \cite{cancellations}. 

\subsection{A pinch of Warped Extra Dimensions}

The rich flavour physics phenomenology of RS type models \cite{Randall:1999ee} 
has been studied numerously. While the basic ideas and features were already established some 
time ago, see e.g.~\cite{Agashe:2004cp}, several involved phenomenological studies of  
flavour observables in RS models with \cite{RScust} and without \cite{RSNocust} 
custodial protection were performed recently. 

For brevity, we refer to Gilad Perez's talk \cite{ckm2010} for the details of 
the setup. The main features affecting the CKM matrix are:
\begin{itemize}
 \item Mixing of the SM fermions with their KK modes: The SM CKM matrix is
       then just a $3\times 3$ sub matrix of the fermion mixing matrix.
       Due to the localisation of the KK wave functions close to the 
       so-called infrared brane, mainly the $t$ and to lesser extent the $b$
       quark will be affected by mixing with KK modes.
 \item As the $W$ also develops KK modes, the zero modes can receive KK mode
       admixtures during EWSB. This will also modify the CKM couplings.
 \item Finally, depending on the way one defines the CKM matrix, the direct 
       effect of the KK modes has to be incorporated\footnote{If the $W$ can go on-shell, 
       like in single-$t$ production, this effect will be strongly suppressed.}.
        E.g.~if one defines the elements via the couplings of effective four-fermion
       operators the whole tower of $W$ modes\footnote{Note that the measurement 
       of $G_F$ via $\mu$ lifetime will not yield the coupling of the SM $W$ 
       (the zero mode), but the effective sum over all modes.} has to be taken 
       into account.
\end{itemize}
%
%
The potential size of the unitarity violation has been studied in detail in \cite{Buras:2009ka} 
for the custodially protected setup and the largest effects are of the order 
$$1 -|V_{tb}|^2 - |V_{ts}|^2 - |V_{td}|^2 \approx \mathcal{O}(5\%)$$
and, as expected, stem from the top quark. Note that in this case the CKM 
matrix is defined via the couplings of the $W$; hence, a straightforward comparison 
with \cite{Buras:2009ka} is not possible, as the minimal RS setup was investigated using the 
effective theory definition of CKM matrix elements. 

\begin{figure}
 \centering
\begin{minipage}{0.45\textwidth}
  \includegraphics[width=0.78\textwidth]{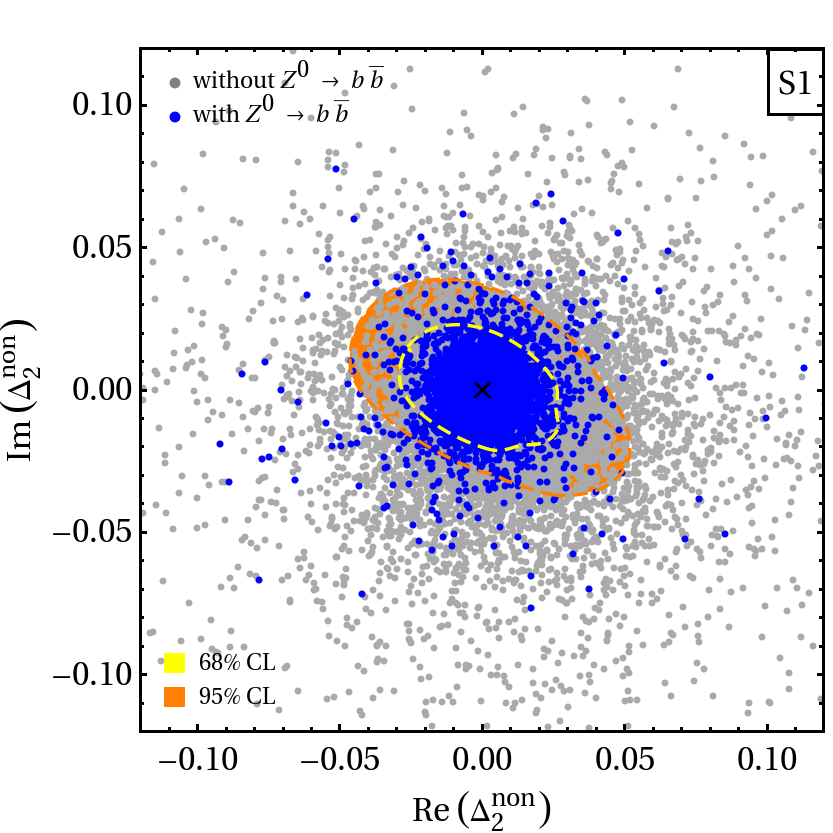}
\end{minipage}
\begin{minipage}{0.45\textwidth}
  \includegraphics[width=0.78\textwidth]{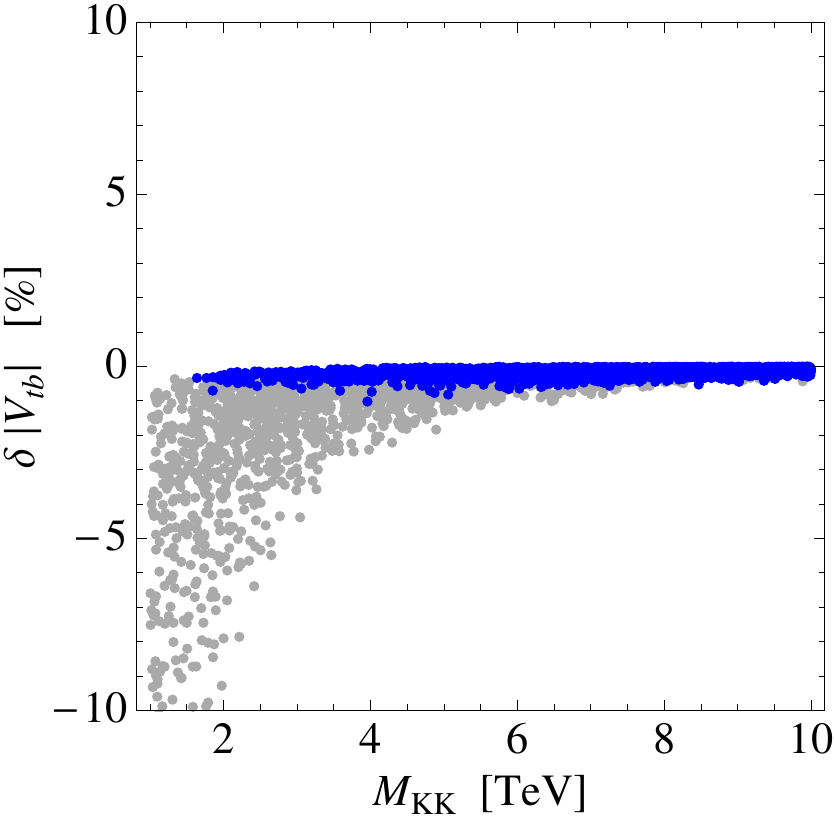}
\end{minipage}
\caption{Unitarity violation in minimal RS model. The left plot show shows the quantity 
$\Delta_2=V_{ud} V^*_{ub}+V_{cd} V^*_{cb}+V_{td} V^*_{tb}$; the
right plot shows values for $(V^{RS}_{tb}-V^{SM}_{tb})/V^{SM}_{tb}$. 
Grey point do not satisfy limits from $Z\to \bar b b$. 
Plots taken from \cite{RSNocust}. \label{fig:rsminimal}}
\end{figure}

As can be seen in Fig.~\ref{fig:rsminimal} the CKM elements 
are in a sense protected by the $Z\to \bar bb$ vertex in the minimal setup; 
this indicates that $\mathcal{O}$(5\%) effects should be expected if a 
custodial symmetry is invoked.

\section{Summary}

Even though the SM values of the CKM matrix elements $V_{td}$ and $V_{ts}$ 
cannot be extracted directly from experiment, unitarity of the CKM matrix alone is 
enough to determine the absolute values with good accuracy. 
An overconstrained fit to the multitude of 
flavour observables, while sensitive to effects of new physics,
does not only provide a powerful self-consistency check of the unitarity condition but
also allows the determination of $V_{td}$, $V_{ts}$ and $V_{tb}$ with 
very high precision.

However, extensions of the standard model with a non-unitray 
mixing matrix for the SM quark are numerous. To provide an idea 
to what magnitude the $V_{tx}$ elements can differ from their SM values
we briefly discussed three models with a non-unitary 'Standard Model 
CKM matrix': vector-like quarks, Fourth Generation models and 
Warped Extra Dimensions.

Each model is capable to survive the bounds set by the various 
experiment and still allows for sizable modification of the CKM 
elements --- especially the 'small' elements $V_{ts}$ and $V_{td}$
can deviate by $\mathcal{O}(50\%)$. Still, if the current experimental 
central value for $V_{tb}$, $0.88$, would be strengthened by LHC and
if theoretical uncertainties could be reduced accordingly all three models
would be hard pressed to accommodate for this. \vspace{-1.0cm}

\Acknowledgements
I would like to thank Ulrich Haisch for illuminating discussions
on the flavour structure of RS models and for providing the 
nice plots.
Furthermore, I thank the organisers of CKM2010 for the excellent work.
This work was supported in part by the DFG SFB/TR 9
``Computergest\"utzte Theoretische Teilchenphysik''.

\end{document}